\documentclass[mathleft
]{an}
\usepackage{graphicx}
\usepackage{times}
\begin{document}

\Pagespan{1}{}
\Yearpublication{}%
\Yearsubmission{}%
\Month{}%
\Volume{}%
\Issue{}%

\title{Exploring the Nature of the Brightest Hyper-luminous X-ray Source}

\author{S. A. Farrell\inst{1}\fnmsep\thanks{Corresponding author:
  \email{saf28@star.le.ac.uk}\newline}
\and  M. Servillat\inst{2}
\and
K. Wiersema\inst{1}
\and
D. Barret\inst{3,4}
\and
O. Godet\inst{3,4}
\and
I. Heywood\inst{5}
\and
T. J. Maccarone\inst{6}
\and
S. R. Oates\inst{7}
\and
B. Plazolles\inst{3,4}
\and
N. A. Webb\inst{3,4}
}
\titlerunning{Exploring the Nature of the Brightest Hyper-luminous X-ray Source}
\authorrunning{S. A. Farrell et al.}
\institute{
Department of Physics and Astronomy, University of Leicester, University Road, Leicester, LE1 7RH, UK
\and
Harvard-Smithsonian Center for Astrophysics, 60 Garden Street, MS-67, Cambridge, MA 02138, USA
\and
Universit\'{e} de Toulouse, UPS, CESR, 9 Avenue du Colonel Roche, F-31028 Toulouse cedex 9, France
\and
CNRS, UMR5187, F-31028 Toulouse, France
\and
Department of Astrophysics, University of Oxford, Keble Road, Oxford, OX1 3RH, UK
\and
School of Physics and Astronomy, University of Southampton, Hampshire, SO17 1BJ, UK
 \and
 Mullard Space Science laboratory/UCL, Holmbury St. Mary, Dorking, Surrey, RH5 6NT, UK
}

\received{30 May 2005}
\accepted{11 Nov 2005}
\publonline{later}


\abstract{%
 The small subset of hyper-luminous X-ray sources with luminosities in excess of $\sim$10$^{41}$ erg s$^{-1}$ are hard to explain without the presence of an intermediate mass black hole, as significantly super-Eddington accretion and/or very small beaming angles are required. The recent discovery of HLX-1, the most luminous object in this class with a record breaking luminosity of $\sim$10$^{42}$ erg s$^{-1}$ in the galaxy ESO 243-49, therefore currently provides some of the strongest evidence for the existence of intermediate mass black holes. HLX-1 is almost an order of magnitude brighter than the other hyper-luminous sources, and appears to exhibit X-ray spectral and flux variability similar to Galactic stellar mass black hole X-ray binaries. In this paper we review the current state of knowledge on this intriguing source and outline the results of multi-wavelength studies from radio to ultra-violet wavelengths, including imaging and spectroscopy of the recently identified optical counterpart obtained with the Very Large Telescope. These results continue to support an intermediate mass black hole in excess of 500 M$_\odot$.}
\maketitle

\section{Introduction}

Ultra-luminous X-ray Sources (ULXs) are extragalactic objects located outside the nucleus of the host galaxy with bolometric luminosities above the Eddington limit for a stellar mass black hole (i.e. $>$ 2.6 $\times$ 10$^{39}$ erg s$^{-1}$ for a 20 M$_\odot$ black hole; see Fabbiano \& White 2006 for a review of ULXs). These extreme luminosities, if the emission is isotropic and below the Eddington limit, imply the presence of a new class of "intermediate mass" accreting black holes with masses between $\sim$10$^2$ -- 10$^5$ M$_\odot$. However, hyper-accretion (Begelman 2002) and/or beaming (Freeland et al. 2006; King 2008) can cause a stellar mass black hole to appear to exceed the Eddington limit. The most luminous ULXs -- the hyper-luminous X-ray sources (HLXs) -- have luminosities in excess of $\sim$10$^{41}$ erg s$^{-1}$, and are more difficult to explain through hyper-accretion or beaming.

The brightest HLX currently known is HLX-1in the edge-on S0a spiral galaxy ESO 243-49, with a derived maximum luminosity (assuming it is at the galaxy redshift of z = 0.0224) of $\sim$10$^{42}$ erg s$^{-1}$ (Farrell et al. 2009). This object is almost an order of magnitude more luminous than the other HLXs (e.g. Gao et al. 2003), and is currently the strongest candidate for hosting an intermediate mass black hole. Since its discovery, a large amount of multi-wavelength archival and new data has been scrutinised in an attempt to uncover the nature of this source. In this paper we will review the current state of knowledge on HLX-1.

\section{X-ray Observations}


HLX-1 was discovered serendipitously while mining the 2XMM catalogue (Watson et al. 2009) for soft spectrum X-ray sources (Farrell et al. 2009). The field of ESO 243-49 was observed in November 2004 for $\sim$22 ks during an observation of the galaxy group centered on the galaxy IC 1633. The  European Photon Imaging Camera (EPIC) pn, MOS1 and MOS2 X-ray spectra during this observation were best fitted with a simple absorbed power-law\footnote{Soria et al. (2010b) have recently argued that the spectrum during this observation is better modeled with the addition of a soft thermal component, citing an F-test significance of $\sim$95\% as justification for the component addition. Disregarding the fact that the quoted F-test probability is only at the 2$\sigma$ level, the use of the F-test in this scenario is seriously flawed. Protassov et al. (2002) have shown that the F-test is invalid when assessing the significance of an addition feature such as a blackbody component. A much more robust approach is to use simulations such as the posterior predictive $p$-values method (Hurkett et al. 2008) used by Farrell et al. (2009) to show a soft thermal component was not statistically required.} with a neutral hydrogen column density of 0.08 $\pm$ 0.03 $\times$ 10$^{22}$ atom cm$^{-2}$ and a photon index of  3.4 $\pm$ 0.3, giving an unabsorbed luminosity in the 0.2 -- 10 keV band of 1.1 $^{+0.01}_{-0.4}$ $\times$ 10$^{42}$ erg s$^{-1}$ assuming it is located within ESO 243-49 (Farrell et al. 2009). 

A follow-up $\sim$50 ks \emph{XMM-Newton} observation was performed in November 2008. Although the flux did not appear to have differed significantly within the errors from the first observation, the spectrum during this second observation was not adequately modeled with a simple absorbed power-law. Instead, the addition of a soft thermal component in the form of a multi-coloured disc blackbody component significantly improved the fit ($\Delta\chi^2$ $\sim$ 160 for 2 fewer degrees of freedom). The X-ray spectrum during this observation was best fitted with an absorbed power-law plus disc blackbody with a neutral hydrogen column density of 0.04 $\pm$ 0.01 $\times$ 10$^{22}$ atom cm$^{-2}$, a photon index of  2.2  $^{+0.4}_{-0.3}$, and an inner disc temperature of 0.18 $\pm$ 0.01 keV, giving an unabsorbed luminosity in the 0.2 -- 10 keV band of 6.4 $\pm$ 0.6 $\times$ 10$^{41}$ erg s$^{-1}$, where the disc blackbody component contributes $\sim$80 $\%$ of the flux (Farrell et al. 2009). Although the luminosity appears to have dropped since the first \emph{XMM-Newton} observation, the values overlap within the errors so we cannot claim flux variability between the two observations, although the spectral shape did change significantly.  



In order to facilitate multi-wavelength studies of HLX-1 and search for an optical counterpart, we first needed to refine the X-ray position to sub-arcsecond accuracy. We thus obtained a 1 ks observation of HLX-1 with the HRC-I camera on \emph{Chandra}  in July 2009. The predicted HRC count rate, if the flux and spectrum remained the same as that observed with \emph{XMM-Newton}, was 0.03 count s$^{-1}$. This should have provided sufficient counts for us to obtain a precise position. However, no source was detected within the {\it XMM-Newton} error circle of HLX-1, indicating the count rate had dropped by at least a factor of 5 (Webb et al. 2010a). \emph{Swift} XRT observations of HLX-1 confirmed the drop in flux and found that one month later the flux increased significantly (Godet et al. 2009). Following this re-brightening we obtained a second deeper 10 ks  observation with the HRC-I  in August 2009, detecting HLX-1 with a net count rate of 0.098 $\pm$ 0.003  cts s$^{-1}$, indicating an increase in flux by a factor of $>$16 between the two \emph{Chandra} observations. After correcting the astrometry by cross-matching detected sources against the 2MASS catalogue, a final position of RA = 01h 10m 28.29s, Dec = -46$^\circ$ 04' 22.3" was obtained for HLX-1, with a 95\% error of 0.3" (Webb et al. 2010a).

\begin{figure}
\includegraphics[width=8cm]{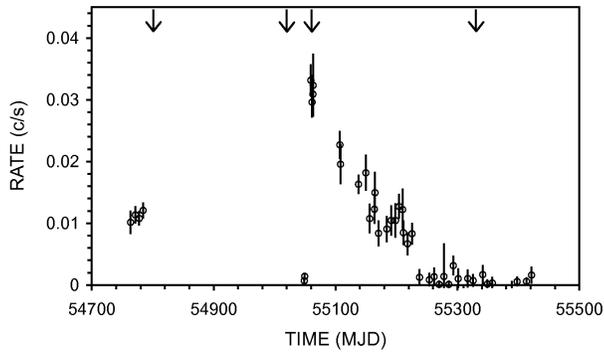}
\caption{\emph{Swift} XRT light curve from the long-term monitoring campaign. The arrows indicate the times of the second \emph{XMM-Newton}, first \emph{Chandra}, second \emph{Chandra}, and third \emph{XMM-Newton} observations from left to right respectively. }
\label{swiftlc}
\end{figure}

Following the re-detection of HLX-1 in August 2009, it has been monitored regularly with \emph{Swift} in X-ray and UV wavelengths. HLX-1 has exhibited dramatic variability over this time, increasing in count rate by a factor of $\sim$40 over $\sim$1 week in August 2009 to 1.1 $\pm$ 0.1 $\times$ 10$^{42}$ erg s$^{-1}$ (Figure \ref{swiftlc}). In this high flux state the spectrum can be best fitted by an absorbed disc blackbody model without the requirement for an additional power law component\footnote{Soria et al. (2010b) derive a slightly lower luminosity during this state using various thermal models with the addition of a weak power law component. However, the $\chi^2$/d.o.f. of 14.4/21 for the pure disc blackbody fit indicates that the statistics are too poor to accurately constrain the spectral shape in this data.} within the limited band pass (and reduced sensitivity) of the XRT (Godet et al. 2009). 

Since the dramatic re-brightening in August 2009, the flux of HLX-1 has been gradually decaying. An XRT observation taken on the 13th of August 2010 measured a count rate of 0.002 $\pm$ 0.001 count s$^{-1}$, consistent with the lowest count rate of 0.0008 $\pm$ 0.0004 count s$^{-1}$ in August 2009. In May 2010 we triggered a 100 ks observation with \emph{XMM-Newton} aimed at constraining the shape of the HLX-1 spectrum during this low flux state (Farrell et al. 2010). Preliminary analysis of this data indicates HLX-1 was at a luminosity of $\sim$3 $\times$ 10$^{40}$ erg s$^{-1}$ during this observation, almost two orders of magnitude below the highest luminosity measurement obtained to date.

\begin{figure}
\includegraphics[width=8cm]{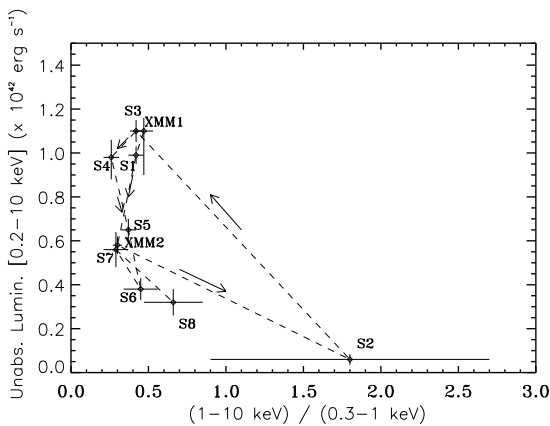}
\caption{Hardness intensity diagram constructed from the \emph{Swift} XRT data (S2 -- S8), with points representing the \emph{XMM-Newton} data from the first (XMM1) and second (XMM2) observations.
 }
\label{hid}
\end{figure}

HLX-1 has been undergoing significant spectral variability in conjunction with the observed large scale flux variability, with the spectrum varying between a state dominated by the thermal component and  a steep power-law state where the thermal component is not present (Godet et al. 2009). The hardness ratios (defined as the ratio of the 1 -- 10 keV count rates over the 0.3 -- 1 keV rates) vary significantly with changing luminosity in a manner reminiscent of the hysteresis behaviour seen in Galactic stellar mass black hole binaries (Figure \ref{hid}; e.g. Maccarone \& Coppi 2003). The hardness ratio errors during the low flux states are large, precluding us from definitively determining using the \emph{Swift} data alone whether HLX-1 is in the canonical low/hard state. However, the preliminary analysis of our most recent \emph{XMM-Newton} observation confirms that the spectrum of HLX-1 was significantly harder than it was during the two previous \emph{XMM-Newton} observations (with a pn hardness ratio of 1.0 $\pm$ 0.2 compared to 0.30 $\pm$ 0.06 and 0.268 $\pm$ 0.007, calculated using the same definition in Godet et al. 2009), with a power-law photon index of $\sim$2 (Farrell et al. 2010). It thus appears that HLX-1 is in a much harder state at lower luminosities than when it is in the high state. However, additional analysis of the timing data is required in order to conclusively determine whether HLX-1 is in the low/hard state (and thus the first ULX to undergo the same spectral hysteresis variability as stellar mass black hole X-ray binaries).

\section{Ultraviolet Observations}

\begin{figure}
\includegraphics[width=8cm]{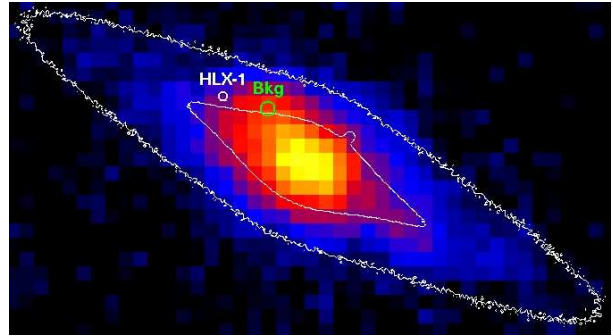}
\caption{\emph{Uvw2} image from 150 ks of \emph{Swift} UVOT data. The white contours show the orientation of the galaxy, and the white circle indicates the \emph{Chandra} position of HLX-1. The green circle indicates the position of the background emission-line source (see Figure \ref{2dspec}).}
\label{uvot}
\end{figure}

In conjunction with the X-ray observations, HLX-1 has been observed regularly with the Ultraviolet and Optical Telescope (UVOT) obtained as part of the \emph{Swift} monitoring campaign, with most of the data obtained in the \emph{uvw2} filter ($\sim$1600 -- 2500 \AA; Webb et al. 2010a). As of the 14th of August 2010 $\sim$150 ks of data had been taken in this filter, providing the deepest and highest resolution UV imaging of this field to date. The core of ESO 243-49 is clearly detected as an extended source in the co-added image from all these observations, with some hints of elongation towards the position of HLX-1 (Figure \ref{uvot}). A similar extension is seen in the \emph{GALEX} near- and far-UV images (Webb et al. 2010a), indicating that this feature is likely to be real. However, with the relatively low resolution of the UVOT and \emph{GALEX} images, it is not clear at this stage whether the emission is truly extended or represents emission from an unresolved point source, or even whether it is associated with HLX-1 or ESO 243-49.

\section{Optical Observations}

The refined X-ray source position led to the discovery of a faint optical counterpart by Soria et al. (2010a) in the $V$-  (24.5 mag) and $R$-bands (23.8 mag) using the Baade Magellan telescope. Assuming this is the true counterpart, the maximum X-ray to optical flux ratio (F$_x$/F$_{opt}$) of HLX-1 is $\sim$1000, two orders of magnitude higher than is typically seen for AGN (e.g. Severgnini et al. 2003), making a background source highly unlikely. However, it has recently been argued that the X-ray spectrum, optical colours\footnote{It should be noted that with magnitudes in only two bands, the optical colours are also consistent with many other types of objects.} and F$_x$/F$_{opt}$ ratio of HLX-1 were consistent with a Galactic quiescent neutron star X-ray binary (Soria et al. 2010a, 2010b). However, the observed large scale variability of HLX-1 is highly unusual for an object of this kind (e.g. Servillat et al. 2008). Nonetheless, in order to determine whether HLX-1 is truly associated with ESO 243-49 we sought optical spectroscopic observations so as to obtain a redshift. 

\begin{figure}
\includegraphics[width=8cm]{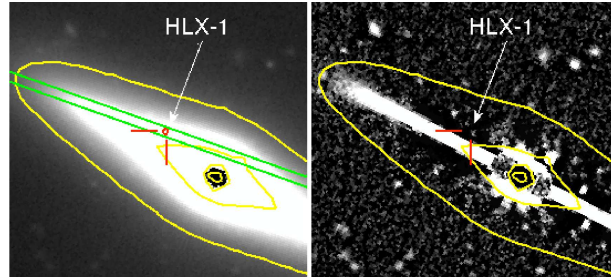}
\caption{\emph{Left:} VLT $I$-band image of ESO 243-49. The green lines indicate the orientation of the slit, and the position of HLX-1 is marked with the red ticks and circle. \emph{Right:} residual image generated by subtracting a 2-D median-filter smoothed image from the un-smoothed image. The contours indicate the orientation of ESO 243-49 taken from the un-smoothed image.}
\label{vltimsub}
\end{figure}

Following the detection of the counterpart we obtained optical spectroscopic data using the Very Large Telescope (VLT; Wiersema et al. 2010). We acquired deep spectroscopy on four nights in November and December 2009, using the FORS2 instrument on the Antu telescope of the VLT with the 300I grism using the OG590 filter. We also performed $\sim$30 min of imaging in the $I$-band (Figure \ref{vltimsub}, left). To confirm the detection of the optical counterpart, we utilised a similar method as Soria et al. (2010a). We smoothed the image along the axis of the galaxy using a 2-D median-filter, and then subtracted the smoothed image from the un-smoothed image so as to remove diffuse emission from the host galaxy. The resulting image shows a clear residual consistent with the \emph{Chandra} position of HLX-1 (Figure \ref{vltimsub}, right), confirming the detection of an optical counterpart. An accurate determination of the $I$-band magnitude is a work in progress (Webb et al. 2010b).

\begin{figure}
\includegraphics[width=8cm]{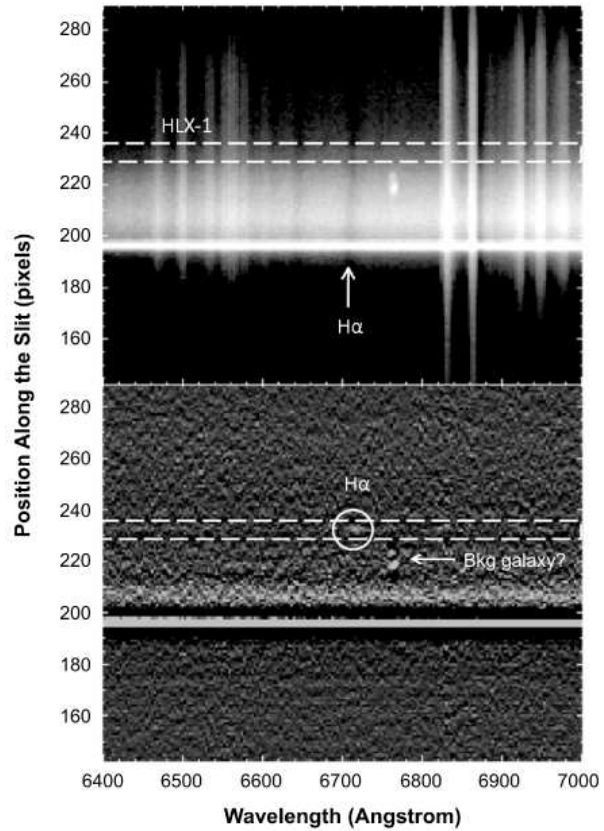}
\caption{\emph{Top:} 2-D spectrum from the VLT FORS2 observations. The H$\alpha$ absorption line at 6716 \AA~ is indicated by the arrow. The bright emission line source just below the HLX-1 trace is likely to be a background galaxy. \emph{Bottom:} Residual spectrum produced by subtracting the smoothed 2-D spectrum from the un-smoothed spectrum. A residual in the HLX-1 trace at the wavelength of H$\alpha$ at the redshift of ESO 243-49 is clearly present. The position of HLX-1 is indicated with the dashed horizontal lines in both panels.}
\label{2dspec}
\end{figure}

The 2-D spectrum shows strong absorption bands typical of a S0 type galaxy to be present in the galaxy bulge (Figure \ref{2dspec}, top). The Na ID and H$\alpha$ absorption features were clearly detected, giving a redshift for the galaxy of z = 0.0223 consistent with previous measurements (Afonso et al. 2005). In addition to these features, a faint trace can be seen at the position of HLX-1, despite its proximity to the bright galaxy nucleus. The H$\alpha$ absorption feature is also present in the HLX-1 trace, and is partially filled in. The bright trace in the lower half of Figure \ref{2dspec} is the point source at RA = 01h10$\arcmin$27.4$\arcsec$, Dec = -46$^\circ$04$\arcmin$25.3$\arcsec$ to the North West of the bulge of ESO 243-49. The absence of the H$\alpha$ absorption line in this trace indicates it is a foreground star. In contrast, the fact that the H$\alpha$ absorption feature is clearly present in the HLX-1 trace indicates it cannot be a foreground object, and must therefore be in or behind ESO 243-49. 

Another interesting feature in Figure \ref{2dspec} is the bright emission line source just below the HLX-1 trace at an approximate position of RA = 01h10$\arcmin$28$\arcsec$, Dec = -46$^\circ$04$\arcmin$23$\arcsec$. The position of this object is $\sim$3$\arcsec$ from the \emph{Chandra} position of HLX-1, well outside the X-ray error circle and therefore clearly not associated with HLX-1. The emission feature appears extended and is clearly resolved into two separate velocity components, possibly indicative of rotation. We therefore conclude that this feature is most likely associated with a background galaxy. If the feature is the H$\alpha$ line, this puts the background galaxy at a redshift of $\sim$0.03. It is interesting to note that the position of this object is also coincident with the extended UV emission in the UVOT \emph{uvw2} image, closer to the core of the extended UV lobe (see Figure \ref{uvot}). It is therefore possible that the UV emission that we speculate could be associated with HLX-1 may instead be linked to this other source.

\begin{figure}
\includegraphics[width=8cm]{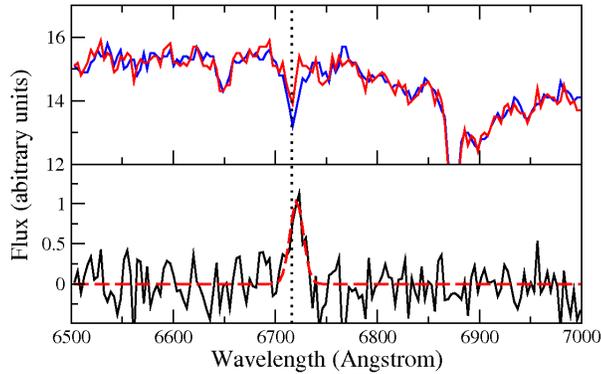}
\caption{\emph{Top:} 1-D spectrum extracted from the position of HLX-1 (red) and a background spectrum interpolated from neighbouring sub-apertures (blue). The absorption line at 6716 \AA~ is H$\alpha$ at the redshift of ESO 243-49. \emph{Bottom:} background subtracted spectrum of HLX-1, showing line emission at 6721 \AA~with a significance of 11.3$\sigma$. The red dashed line indicates the Gaussian model fitted to the emission line. The dotted vertical line indicates the wavelength of the H$\alpha$ absorption line in the ESO 243-49 spectrum.}
\label{vltspec}
\end{figure}

The HLX-1 background subtracted spectrum reveals an emission line with a significance of 11.3$\sigma$ and an approximately Gaussian profile superimposed on a very weak continuum (Figure \ref{vltspec}). The central wavelength of this emission line is 6721 \AA, consistent with it being H$\alpha$ at a redshift of z = 0.0223. The luminosity of this line is $\sim$10$^{37}$ erg s$^{-1}$, although the errors introduced through the background subtraction prohibit us from estimating a luminosity for the continuum emission. The offset of this line from the ESO 243-49 redshift is $\sim$170 km s$^{-1}$, considerably smaller than the galaxy rotation curve and therefore consistent with HLX-1 being gravitationally bound to ESO 243-49. The only other plausible identification of this line is [OII] for a source at z = 0.80; however, the F$_x$/F$_{opt}$ ratio of $\sim$1000 argues strongly against a background galaxy. We therefore conclude that the line is most likely H$\alpha$ emission from an object gravitationally bound to ESO 243-49 (Wiersema et al. 2010).   


\section{Discussion}

The detection of the H$\alpha$ emission line at a redshift consistent with that of ESO 243-49 conclusively puts HLX-1 at a distance of $\sim$95 Mpc, thereby ruling out a background AGN or a foreground object such as a Galactic quiescent neutron star X-ray binary. This redshift measurement also confirms HLX-1 as the most extreme ULX with a maximum luminosity of $\sim$10$^{42}$ erg s$^{-1}$. The precise nature of HLX-1, however, is still unclear.

The large scale X-ray variability of HLX-1 rules out a collection of unresolved lower-luminosity sources. A luminous supernova remnant is also ruled out as such an object would be expected to be located in areas where star formation has occurred, which is difficult to reconcile with the location of HLX-1 outside the galaxy plane. We are therefore left with an object within the bounds of ESO 243-49 that appears to exceed the Eddington limit for a stellar mass black hole by a significant margin. The X-ray spectra of HLX-1 observed over a range of luminosities are all consistent with an accreting black hole (e.g. Maccarone 2003), and it is difficult to envisage the X-ray emission arising from any other type of object. The remaining question is whether the black hole in question is stellar mass (presumably undergoing hyper-accretion and/or beaming) or in the intermediate mass range.   

Stellar mass black hole binaries can theoretically exceed the Eddington limit for short periods of time by up to a factor of 10 (Begelman 2002), leading Farrell et al. (2009) to place a conservative lower limit of 500 M$_\odot$ on the mass of the black hole. The spectral variability observed from HLX-1 is entirely consistent with the behaviour of Galactic stellar mass black hole binaries, with the most recent \emph{XMM-Newton} observations confirming that at a luminosity of $\sim$3 $\times$ 10$^{40}$ erg s$^{-1}$ it has a significantly harder spectrum than that observed at luminosities $>$10$^{41}$ erg s$^{-1}$ (compatible with it being in the canonical low/hard state). Stellar mass black hole binaries that are observed at intermediate inclination angles (i.e. with 0$^\circ \gg i \ll 90^\circ$) typically have luminosities of $\sim$1 -- 3$\%$ of the Eddington value in the low/hard state (Maccarone 2003), implying a mass of $\sim$8,000 -- 23,000 M$_\odot$ for HLX-1 if it is in a similar spectral state at its lower luminosity and assuming that (like the Galactic binaries) the observed emission is not beamed towards us. 

It has been suggested that ULXs may represent extragalactic analogues of the Galactic microquasar SS 433 that are viewed ``face-on", such that mild geometric beaming effects and high mass accretion rates could easily explain the apparent super-Eddington luminosities (e.g. King et al. 2001; Begelman, King, $\&$ Pringle 2006). Such systems would thus represent a short-lived yet extremely common stage in the evolution of high mass X-ray binaries (HMXBs), where sufficient thermal-timescale mass transfer is provided by Roche-Lobe overflow accretion from a high mass companion star (King et al. 2001). Support for this theory has been provided by the apparent excess of ULXs found near star forming regions (e.g. Swartz, Tennant $\&$ Soria 2009), where HMXBs should be plentiful. However, it should be noted that while many ULXs are found \emph{near} star forming regions, none have as yet been found within massive star clusters (e.g. Swartz et al. 2009). This is not necessarily a problem, however, as ULXs may be kicked out of the star clusters through anisotropic supernovae explosions\footnote{An alternative possibility is that some ULXs may not be the product of recent star formation episodes, but instead may be the cause. This is a natural conclusion if ULXs represent the nuclei of accreted satellite galaxies, which should drive star formation as they interact with the larger host galaxies.} (A. R. King 2010, private communication). 
HLX-1 is located in the outskirts of ESO 243-49 at $\sim$1 kpc above the plane, far removed from any regions of star formation where a population of HMXBs might be found, although a not unreasonable kick velocity of $\sim$100 km s$^{-1}$ could account for this offset from the plane within the expected lifetime of a high mass companion. 

Relativistic Doppler beaming from a face-on low mass X-ray binary still remains a possibility, although it is difficult to reconcile the observed variability with this model. As noted earlier, HLX-1 appears to follow a similar track in the hardness-intensity diagrams as the Galactic stellar mass black hole binaries that \emph{are not viewed down the jet axis}. In stellar mass black hole systems jets have been observed to turn on during the low/hard state, and turn off following a transition to the high/soft state (e.g. Belloni 2010). It therefore follows that relativistic beaming is likely to be strongest when HLX-1 is at the lower luminosities, with the jets pointing away from us. 

Alternatively, if we are looking at HLX-1 face-on (i.e. down the jet axis), then how do we explain the observed variability? If the jets turn off during the low flux state, then the observed luminosity of $\sim$10$^{40}$ erg s$^{-1}$ it still far too high for a stellar mass black hole. An alternative possibility is that the variability could be tied to a changing viewing angle, in turn linked to the precession of jets through radiation induced warping such as is seen with SS 433 (e.g. Fabrika 2004). Continued long term monitoring in X-rays with the \emph{Swift} observatory is necessary to test this hypothesis, although it seems much more likely that like the Galactic stellar mass black hole binary systems that it appears to resemble, the emission of HLX-1 is not beamed directly towards us. 

Relativistic jets should also produce radio emission with a luminosity that is a function of both the X-ray luminosity and the black hole mass (the black hole "fundamental plane" relation; e.g. K\"{o}rding, Falcke $\&$ Corbel 2006). The deepest radio observation to date of this source has been with the Australia Telescope Compact Array at 1.4 GHz (Hopkins et al. 2003) . The centre of ESO 243-49 was clearly detected, but no emission was evident at the position of HLX-1 down to a 3$\sigma$ upper limit of 45 $\mu$Jy (Webb et al. 2010b). The non-detection is consistent with the black hole fundamental plane relation if HLX-1 is an intermediate mass black hole of $\leq$10$^5$ M$_\odot$, as the expected flux density would be below the sensitivity limits of the most powerful radio telescopes currently available. Alternatively, if the black hole mass was $\geq$10$^6$ M$_\odot$, we would have expected to have detected it with a flux density of $\sim$185 $\mu$Jy, therefore ruling out a super-massive black hole for HLX-1. We would, however, only expect to see radio emission when in the low/hard state, so the non-detection is also consistent with a non-jet state. 

The detection of the H$\alpha$ emission line allows us to place some constraints on the nature of the X-ray emission. The ratio of H$\alpha$ to X-ray luminosity (L$_{H\alpha}$/L$_X$) is $\sim$10$^{-5}$, assuming L$_X$ = 10$^{42}$ erg s$^{-1}$.  
The typical disk photons from HLX-1 will be $\sim250$ times as energetic 
as H$\alpha$ photons. The cross-section for absorption of $\sim0.5$ keV 
electrons is roughly the Thompson cross-section (Bell \& Kingston 1967).  Assuming that the local absorbing column near the 
source is of order or less than the foreground $N_H$ of $4\times10^{20}$ 
cm$^{-2}$, then only $\sim3\times10^{-4}$ of the X-ray luminosity should 
be absorbed.  The re-emission as H$\alpha$ will then produce a luminosity 
of $\sim10^{-6}$ of the X-ray luminosity -- an order of magnitude below that 
observed.  Since the emission line region is likely to be optically thin, 
and hence isotropic, it is unlikely that the X-ray emission is strongly 
beamed toward us, unless the H$\alpha$ emission is primarily due to 
collisional excitation rather than photoionisation.  Given the source 
location far from the galactic plane of ESO 243-49, this seems unlikely.  Additional emission line diagnostics would nonetheless be extremely 
helpful for confirming the usefuless of the H$\alpha$ luminosity as a 
constraint on beaming.

A more likely scenario perhaps is that HLX-1 is the nucleated remnant of an accreted satellite dwarf galaxy. Such a scenario has previously been suggested as a possible origin for the brightest ULXs (King \& Dehnen 2005), and can easily account for the extreme luminosity (assuming that the dwarf galaxy in question contained a central black hole of intermediate mass) and the location of HLX-1 outside the plane of ESO 243-49 (Bellovary 2010). Such a nucleated dwarf galaxy would appear in many respects very similar to a globular cluster, another possible environment in which an intermediate mass black hole might form (e.g. Miller \& Hamilton 2002).  In order to confirm this hypothesis, it is necessary to obtain further deep observations in near-infrared, optical and UV wavelengths so as to construct a spectral energy distribution and thus determine the nature of the environment around HLX-1. To this end, we have been awarded high resolution imaging in near-infrared to far-UV bands with the \emph{Hubble Space Telescope} in cycle 18, which should enable us to shed light on the nature of HLX-1 in the near future.

\acknowledgements

We thank the anonymous referee for their helpful comments. S.A.F., K.W., T.J.M., I.H., and S.R.O. acknowledge STFC funding. T.J.M. thanks the European Union FP7 for support through grant 215212 ÒBlack Hole UniverseÓ. M.S. is supported in part by \emph{Chandra} grants AR9-0013X and GO9-0102X.



\begin{thebibliography}{}
  
  
  \bibitem{} Afonso, J., Georgakakis, A., Almeida, C., Hopkins, A.~M., Cram, L.~E., Mobasher, B., Sullivan, M.: 2005, \apj 624, 135  
  \bibitem{} Bell, K. L., Kingston, A. E.: 1967, MNRAS,136, 241 
  \bibitem{} Belloni, T. M.: 2010, AIP Conf. Proc. 1248, 107
  \bibitem{} Bellovary, J. M., Governato, F., Quinn, T. R., Wadsley, J., Shen, S., Volonteri, M.: 2010, ApJ, 721, L148
  \bibitem{} Begelman, M. C.: 2002, \apj 568, L97
  \bibitem{} Begelman, M. C., King, A. R., Pringle, J. E.: 2006, MNRAS 370, 399
  \bibitem{} Fabbiano, G., White, N. W.: 2006, Compact Stellar X-ray Sources, 475
  \bibitem{} Fabirka, S.: 2004, ASPRv 12, 1
  \bibitem{} Farrell, S. A., Webb, N. A., Barret, D., Godet, O., Rodrigues, J. M.: 2009, Nat 460, 73
  \bibitem{} Farrell, S. A., Webb, N. A., Barret, D., Servillat, M., Godet, O.: 2010, MNRAS, in preparation
  \bibitem{} Freeland, M., Kuncic, Z., Soria, R., Bicknell, G. V.: 2006, \mnras 372, 630
  \bibitem{} Gao, Y., Wang, D. Q., Appleton, P. N., Lucas, R. A.: 2003, \apj 596, L171
  \bibitem{} Godet, O., Barret, D., Webb, N. A., Farrell, S. A., Gehrels, N.: 2009, \apj 705, L109
  \bibitem{} Hopkins, A. M., Afonso, J., Chan, B., Cram, L. E., Georgakakis, A., Mobasher, B.: 2003, AJ, 125, 465
  \bibitem{} Hurkett, C. P., Vaughan, S., Osborne, J. P. et al.: 2008, ApJ 679, 587
  \bibitem{} King, A. R.: 2008, \mnras 385, L113
  \bibitem{} King, A. R., Davies, M. B., Ward, M. J., Fabbiano, G., Elvis, M.: 2001, ApJ, 552, L109
  \bibitem{} King, A. R., Dehnen, W.: 2005, \mnras 357, 275
  \bibitem{} K{\"o}rding, E., Falcke, H., Corbel, S.: 2006, A\&A 456, 439 
  \bibitem{} Maccarone, T. J.: 2003, A\&A 409, 697
  \bibitem{} Maccarone, T. J., Coppi, P. S.: 2003, MNRAS 338, 189
  \bibitem{} Miller, M. C, Hamilton, D. P.: 2002, MNRAS 330, 232
  \bibitem{} Protassov, R., van Dyk, D. A., Connors, A., Kashyap, V. L., Siemiginowska, A.: 2002, ApJ 571, 545
  \bibitem{} Servillat, M., Dieball, A., Webb, N. A. et al.: 2008, A\&A 490, 641 
  \bibitem{} Severgnini, P., Caccianiga, A., Braito, V. et al.: 2003, A\&A 406, 483 
  \bibitem{} Soria, R., Hau, G. K. T., Graham, A. W., Kong, A. K. H., Kuin, N.~P.~M., Li, I.-H., Liu, J.-F., Wu, K.: 2010a, \mnras 405, 870
  \bibitem{} Soria, R., Zampieri, L., Zane, S., Wu, K.: 2010b, \mnras, in press (arxiv:1008.3382)
  \bibitem{} Swartz, D. A., Tennant, A. F., Soria, R.: 2009, ApJ 703, 159
  \bibitem{} Watson, M. G., Schr{\"o}der, A., Fyfe, D. et al.: 2009, A\&A 493, 339
  \bibitem{} Webb, N. A., Barret, D., Farrell, S. A., Godet, O., Heywood, I., Oates, S., Pancrazi, B., Servillat, M.: 2010, ApJ, in preparation 
   \bibitem{} Webb, N. A., Barret, D., Godet, O., Servillat, M., Farrell, S. A., Oates, S. R.: 2010, \apj 712, L107
   \bibitem{} Wiersema, K., Farrell, S. A., Webb, N. A., Servillat, M., Maccarone, T. J., Barret, D., Godet, O.: 2010, ApJ, 721, L102  
 




\end{thebibliography}
\end{document}